\documentclass{pasj00}

\usepackage{multirow}
\usepackage{comment}
\usepackage{graphicx}

\draft

\begin{document}

\SetRunningHead{M. Popov, V. Soglasnov et al}{Giant Pulses from the Crab Pulsar}
\Received{$\langle$$\rangle$}
\Accepted{$\langle$$\rangle$}
\Published{$\langle$$\rangle$}

\title{Multifrequency Study of Giant Radio Pulses from the Crab Pulsar
with the K5 VLBI Recording Terminal}

\author{
Mikhail \textsc{Popov}\altaffilmark{1},
Vladimir \textsc{Soglasnov}\altaffilmark{1}, 
Vladislav \textsc{Kondratiev}\altaffilmark{1,2,3}, 
Anna \textsc{Bilous}\altaffilmark{1,4}, 
Olga \textsc{Moshkina}\altaffilmark{1,4}, 
Vasily \textsc{Oreshko}\altaffilmark{5}, 
Yury \textsc{Ilyasov}\altaffilmark{5}, 
Mamoru \textsc{Sekido}\altaffilmark{6}, \and 
Tetsuro \textsc{Kondo}\altaffilmark{6}
}

\altaffiltext{1}{Astro Space Center of Lebedev Physical Institute, Moscow, Russia}
\email{mpopov@asc.rssi.ru, vsoglasn@asc.rssi.ru}
\altaffiltext{2}{West Virginia University, Department of Physics, Morgantown, USA}
\altaffiltext{3}{National Radio Astronomy Observatory, USA}
\email{vlad.kondratiev@gmail.com}
\altaffiltext{4}{Moscow Institute of Physics and Technology, Russia}
\email{hanna.bilous@gmail.com, MoshkinaOlga@yandex.ru}
\altaffiltext{5}{Puschino Radio Astronomical Observatory, Puschino, Russia}
\email{oreshko@prao.ru, ilyasov@prao.ru}
\altaffiltext{6}{Kashima Space Research Center, NICT, Japan}
\email{sekido@nict.go.jp, kondo@nict.go.jp}

\maketitle

\begin{abstract}

Simultaneous multifrequency observations of the Crab pulsar giant pulses (GPs)
were performed with the 64-m Kalyazin radio telescope at four frequencies 
$\nu =$~0.6, 1.4, 2.2 and 8.3~GHz, using the K5 VLBI recording terminal.
The K5 terminal provided continuous recording in 16 4-MHz wide frequency channels 
distributed over 4 frequency bands.
Several thousand GPs were detected during about 6 hours of observations 
in two successive days in July 2005.
Radio spectra of single GPs were analysed at separate frequencies 
and over the whole frequency range. These spectra manifest notable modulation 
over frequency ranges, $\Delta\nu$, both on 
large ($\Delta\nu/\nu\approx 0.5$) and small ($\Delta\nu/\nu\approx 0.01$)
frequency scales. Cross-correlation analysis of GPs at 2.2~GHz showed that
their pulse shapes can be interpreted as an ensemble 
of unresolved bursts grouped together at time scales of $\approx 1~\mu$s being
well-correlated over a 60-MHz band. The corresponding GP cross-correlation functions
do not obey the predictions of the amplitude-modulated noise model of \citet{Rickett75}, thus
indicating that unresolved components represent a small number of 
elementary emitters.

\end{abstract}

\section{Introduction}
Giant pulses (GPs) represent an extraordinary phenomenon in pulsar radio emission.
First, they have huge peak flux densities, that may be as great as
several millions of janskys \citep{cordes2004,sog2006}. Second,
their duration is ultra short with GP fundamental components being even as short as $<2$~ns
\citep{hankins2003,hanketal2007}. Third, these elementary nanopulses are strikingly almost completely 
circularly polarized, either left- or right-handed \citep{cognard1996,popov2004,hankins2003}.
Giant pulses are reliably known only from seven pulsars 
(see, e.g., \cite[and references therein]{knight2006}),
but the properties of GPs were studied extensively only for
two objects: for the original millisecond pulsar B1937+21 
\citep{cognard1996, kink2000, Sogetal2004,kondratiev2006}, 
and for the
Crab pulsar (see the most recent review by  \cite{bilous2007} and references therein).
In contrast to regular radio pulses,  which typically have Gaussian distribution,
the cumulative distribution of GP energies
follows a power law $N(E>E_0)\propto E_0^{-\alpha}$
with the exponent $\alpha$ being in the range of 1.5--2.5 (\cite{PopStap2007,Bhat2008}).
These authors also reported about
the deviation from the single
power law fit of cumulative distributions at low energies.
No deviations, however, were found at the highest energies even during 160-hours
monitoring of the Crab pulsar with the Kalyazin 64-m radio telescope  at
frequencies of 0.6, 1.65 and 4.85~GHz (\cite{Popetal2008}). These
three-frequency observations demonstrated that the pulse energy, in general,
grows with decreasing frequency, and the spectral index is of
about $-1.8$. But radio spectra of individual giant pulses manifest
deep modulations both at small and large frequency scales
(\cite{hanketal2007,PopStap2004,Popetal2008}).
In this paper we present an analysis of results obtained from simultaneous
4-frequency observations of GPs from the Crab pulsar conducted with the 64-m Kalyazin
radio telescope at the frequencies of 0.6, 1.4, 2.2, and 8.3~GHz using the K5 VLBI recording terminal.
The description of the observing setup and data reduction pipeline is given in  
Section~\ref{sect:obs} and \ref{sect:reduction}. The results of our GP studies at different
frequencies are reported in Sections~\ref{sect:res8.3}, \ref{sect:res1.4-2.2}, and \ref{sect:f600MHz},
and the conclusions are given in  Section~\ref{sect:concl}.


\section{Observations}\label{sect:obs}

\begin{figure*}[htb]
\begin{center}
\FigureFile(170mm,120mm){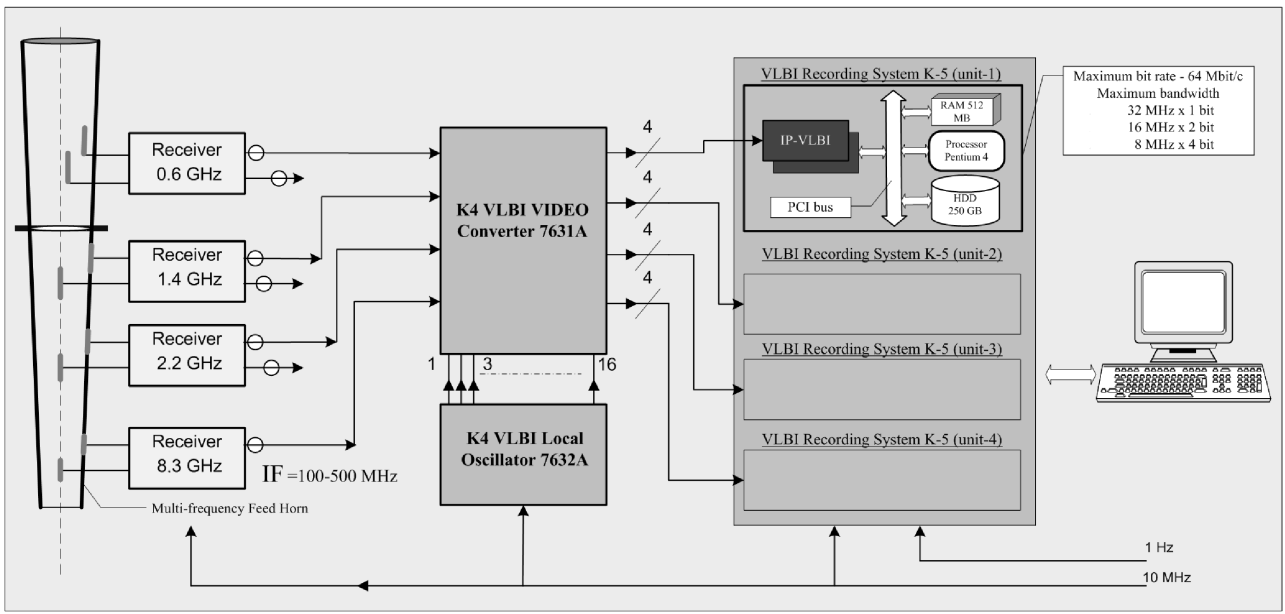}
\end{center}
\caption{Block diagram of antenna feed line of the 64-m Kalyazin radio 
telescope with the K5/VSSP VLBI recording terminal.}\label{block}
\end{figure*}

Multifrequency observations of the Crab pulsar were conducted on July 20--21, 2005 with
the 64-m Kalyazin radio telescope simultaneously at four frequency bands: 0.6, 1.4, 2.2,
and 8.3~GHz. The total observing time was about 6 hours. The simultaneity of observations
at four frequencies was possible owing to the multifrequency feed that the Kalyazin radio telescope
is equipped with. To record the data we used the K5/VSSP VLBI recording terminal (\cite{Kondo2002})
and the K4 baseband converter provided by the  Kashima VLBI group of the National
Institute of Information and Communications Technology (NICT, Japan) for
joint VLBI observations of pulsars under mutual agreement. The block diagram
of the antenna feed line with the K5/VSSP VLBI recording terminal is presented in Figure~\ref{block}.
The terminal consists of 4 workstations with K5/VSSP PCI-bus boards, 
that allow to record up to 16 channels, each with
the bandwidth of 4~MHz. The channels may be distributed
 in frequencies and/or polarizations by using the K4 baseband converter.
In our observations we used the frequency setup
presented in  Table \ref{tab:freq}.
the channel frequencies $f_1$--$f_4$ from the Table correspond to the lower edge of each frequency channel,
and only upper sub-bands were recorded. 
At the frequency band of 600~MHz only one 4-MHz
channel was used because of limitations by radio frequency interference.
In all other bands we used four channels distributed along
the whole band as indicated in  Table~\ref{tab:freq}.

\begin{table*}
\caption{Observations summary. Columns list the frequency band, the frequency channels setup for every band 
with the values corresponded to the lower edge of the channel, $f_1$--$f_4$, the system equivalent flux density (SEFD) including the contribution from the Crab nebula (see text), the estimated dispersion smearing time in a single frequency channel, $\tau_\mathrm{DM}$, 
the total number of events found in a total band, $N_\mathrm{p}$, the peak flux density of the strongest event, $S_\mathrm{max}$.}\label{tab:freq}
  \begin{center}
    \begin{tabular}{ccccccccc}
      \hline
      \multirow{2}{*}{Band} & \multicolumn{4}{c}{Frequency channels (MHz)} & SEFD &$\tau_\mathrm{DM}$ &\multirow{2}{*}{${N_\mathrm{p}}^{*}$} & $S_\mathrm{max}$ \\
\cline{2-5}
           & $f_1$ & $f_2$ & $f_3$ & $f_4$ & (Jy) & ($\mu$s) &  & (MJy) \\
      \hline
      600 & 599.6 & -- & -- & -- & 1300& 8500 &$\sim10000^{**}$& 2.2 \\
      1400& 1395.0& 1410.0& 1425.0& 1440.0& 1150&650&1112&3.7\\
      2200& 2140.0& 2160.0& 2180.0& 2200.0& 1050&180&631&1.3\\
      8300& 8260.0& 8280.0& 8300.0& 8320.0& 240& 3&1&0.04\\
      \hline
      \multicolumn{9}{l}{${}^{*}$ \small Giant pulses can be simultaneously detected in more than one frequency channel,} \\
      \multicolumn{9}{l}{\small i.e. the actual number of unique pulses detected is smaller than the total number of} \\
      \multicolumn{9}{l}{\small events in all frequency channels, $N_\mathrm{p}$.}\\
      \multicolumn{9}{l}{${}^{**}$ \small See section \ref{sect:f600MHz} for explanation.}
    \end{tabular}
  \end{center}
\end{table*}

The system equivalent flux density (SEFD) for every band is dominated by the contribution
from the Crab nebula and is given in the Table~\ref{tab:freq}.
Flux densities for the Crab nebula were calculated by using the relation $S_{f}=955 f^{-0.27}$,
with $f$ expressed in GHz (\cite{cordes2004}). 
At 8.3~GHz solid angle of the radio telescope beam covers
only 20\% of the area occupied by the nebula. Thus, we estimated the contribution from the Crab nebula to be 
5 times less at this frequency assuming an uniform surface brightness distribution.

\section{Data reduction}\label{sect:reduction}

The core of the processing pipeline consisted of successive decoding of the data, correction for
bit-statistics  
and dedispersion. Every single frequency channel was decoded according to  the K5/VSSP (IP-VLBI)
Board Data format\footnote{{\tt http://www2.nict.go.jp/w/w114/stsi/K5/VSSP/data\_format\_e.pdf}} to
produce the raw binary data stream. The decoded data represented the stream of only four alternating
numbers: $-3$, $-1$, $+1$, and $+3$ according to the VLBA coding of 2-bit digitized values. 
These data were corrected then for the bit-statistics, i.e. the decoded values were adjusted to real ones
that correspond to instantaneous $\pm1\sigma$ brightness levels by using the technique developed by \citet{jenet1998}.
To coherenetly dedisperse the data we used the predetection dispersion removal technique
pioneered by Hankins (1971, see also Hankins \& Rickett, 1975). More details about the correction for
bit-statistics and coherent dedispersion can be found in \citet{Popetal2002}. However, for the
coherent dedispersion, unlike the Eq.~1 from \citet{Popetal2002}, we have used the exact 
equations for phase correction: 
$$
\delta\phi (f)= \frac{2\pi \mathrm{DM}}{Df}{\left(\frac{\Delta f}{f_0}\right)}^2~,
$$
where DM is the dispersion measure in pc cm$^{-3}$, $D = 2.41\times 10^{-16}$ pc cm$^{-3}$ s the 
dispersion constant, $f_0$ the lowest frequency of the frequency channel, 
$f$ the current frequency, and $\Delta f = f - f_0$. 

To allow for the direct comparison of times
of arrival (TOA) of giant pulses, we reduced TOAs in all frequency channels within a band to
the highest frequency (except for the data at 0.6~GHz where we had only one frequency channel).
This was done by introducing a time delay via a linear phase correction $k_f$ in the corresponding spectra
simultaneously with the correction for $\delta\phi (f)$. The linear correction $k_f$ is given 
by $k_f = 2\pi\times dt/N$, where $dt$ is the time delay between corresponding frequencies in samples 
(including fractional part), and $N$ is equal to the number of samples in the array used for the 
Fast Fourier Transform (FFT). We used $N=2097152$, corresponding to the time interval of 262.144~ms and 
covering about 7.8 periods of the Crab pulsar. The dedispersion correction of every such chunk
of data resulted in a number of spoiled samples in the restored signal due to the cycle nature
of the convolution made through the FFT. The duration of this spoiled portion is equal to the dispersion
time in a single frequency channel, $\tau_\mathrm{DM}$, of which the estimated approximate values  are given
in Table~\ref{tab:freq}. To reconstruct the signal without gaps the successive chuncks of data 
were overlapped by $\tau_{DM}$. The used value of DM  $56.738 cm^{-3}pc$ 
was taken from the Jodrell Bank Crab Pulsar Monthly 
Ephemeris\footnote{{\tt http://www.jb.man.ac.uk/\~~$\!\!\!$pulsar/crab.html}} \citep{JBE} 
for the epoch close to our observations.

To search for GPs at all frequencies except 600~MHz\footnote{At 600 MHz 
we did not selected individual GPs (except for the preliminary search). Instead of this, 
we compared the overall statistical properties of signal on emission 
longitude windows with those on the noise.}, we set up the detection threshold at a 
level of $20\sigma$ above the mean value 
for the signal with the original sample tome of 125~ns. The statistics of amplitudes $x$ of such a signal
is close to a $\chi ^2$-distribution with 2 degrees of freedom $P(x>a)=\exp(-a)$, where
$P$ is the probability of the amplitude $x$ of the signal exceeding the threshold $a$,
and $a$ is in units of $\sigma$. 
Thus, the probability of false detection with  a threshold of of $20\sigma$ above the mean 
level is $7.6\times 10^{-10}$, and for the 
sampling rate of 8~MHz it will lead to a detection of one false GP in every 3 minutes.
During the subsequent analysis only events that occurred in the narrow windows located at the longitudes
of the main pulse and the interpulse were selected. These windows
were each about $100\mu$s wide and occupied only 0.6\% of the period. Therefore, false detection 
happened only once per several hours. 

\begin{figure*}
  \begin{center}
    \FigureFile(80mm,80mm){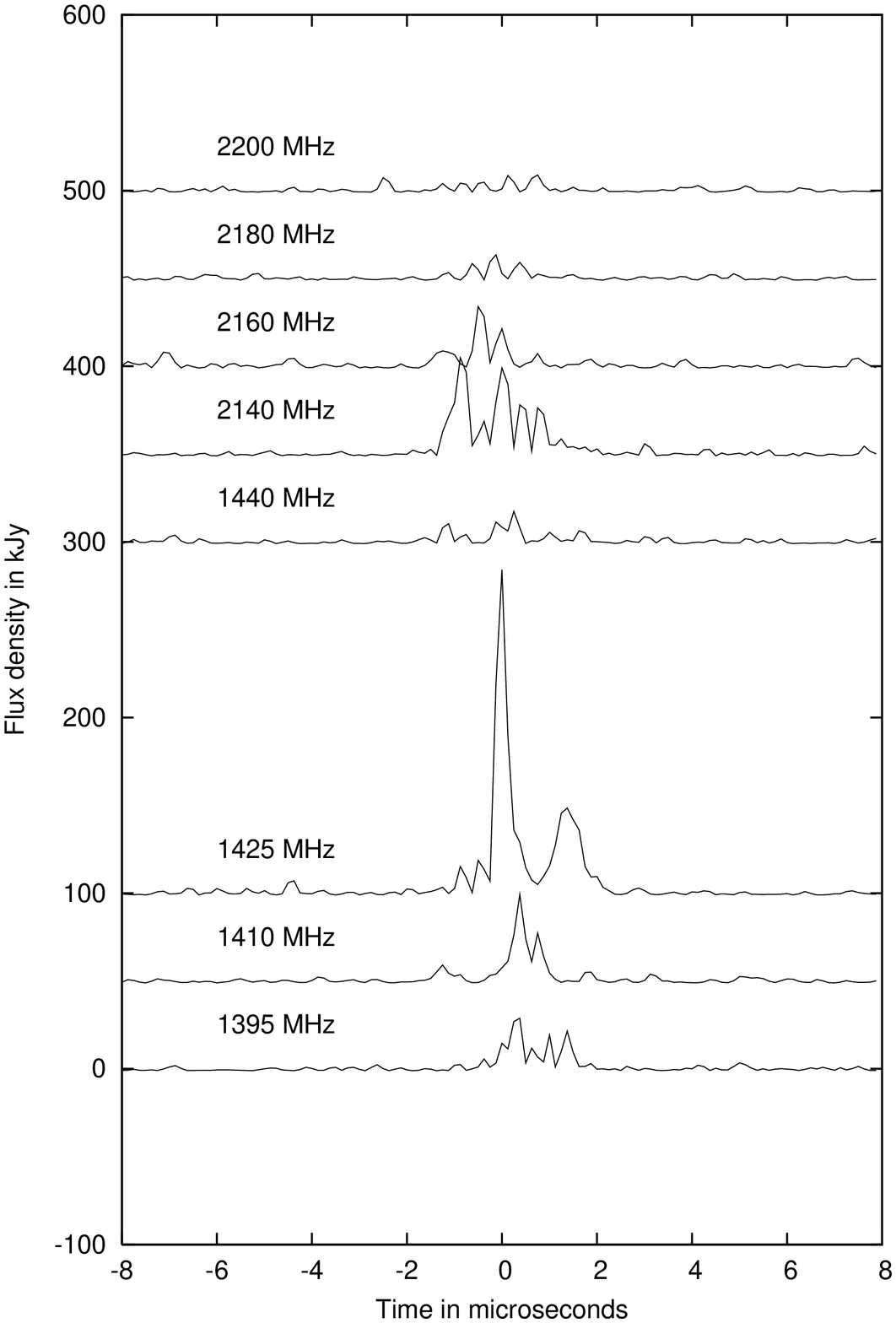}
    \FigureFile(80mm,80mm){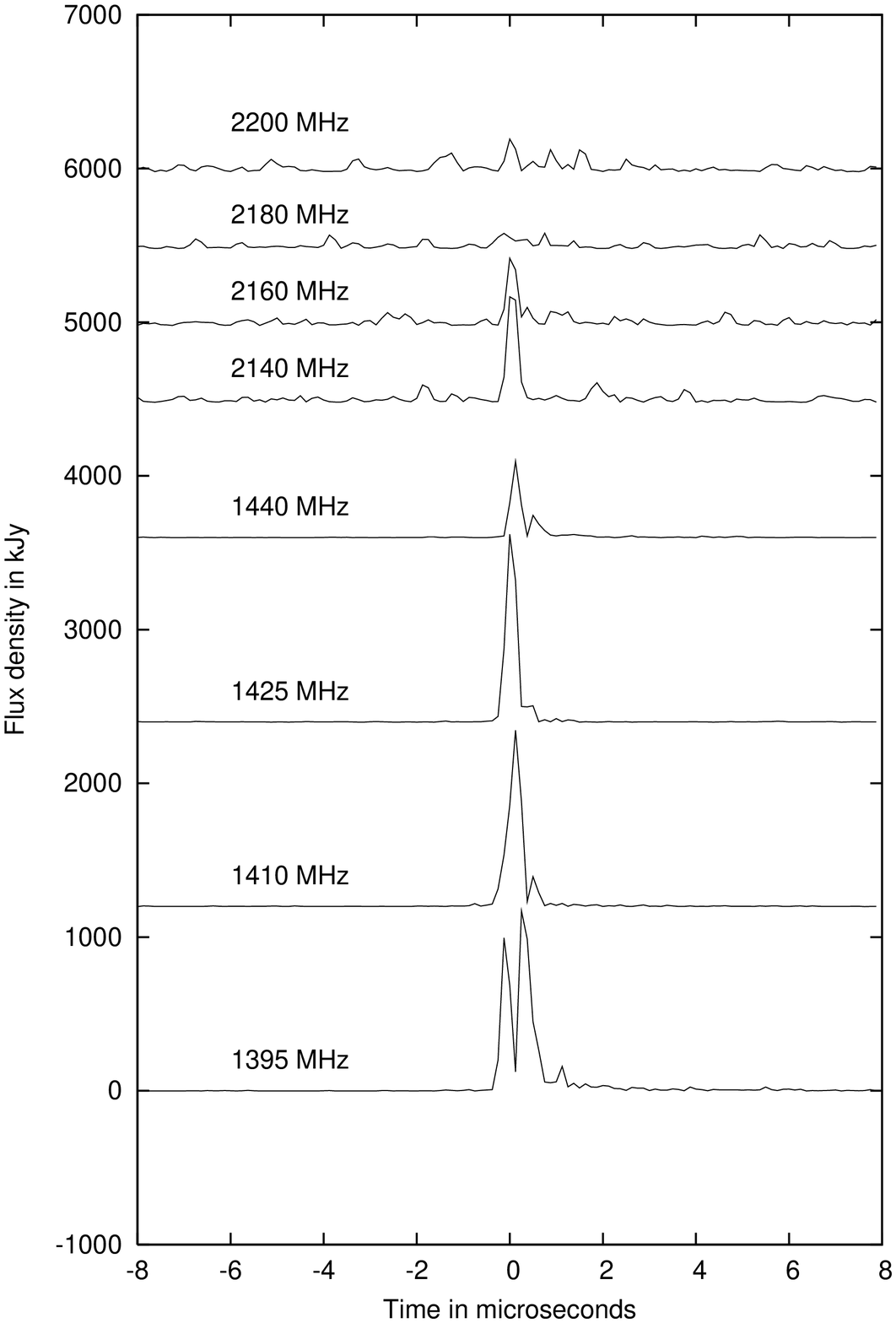}
  \end{center}
  \caption{Example of strong giant pulses detected simultaneously at 1.4 and 2.2~GHz in several frequency
channels. The zero value in time corresponds to the sample with the maximum flux density of the giant pulse
in any of the channels.}
\label{fig:example}
\end{figure*}

The total number of detected events was about 10000, 2700, 1600, and 1 at the frequencies of 0.6, 1.4, 2.2, and 8.3~GHz, 
respectively (see Table~\ref{tab:freq}). 
At 1.4 and 2.2~GHz sufficiently 
strong GPs were detected simultaneously in several frequency channels, and the number
of real GPs is 2--3 times less than the total number of detections.
The brightest GP was found at 1.4~GHz with the peak flux density of 3.7~MJy. Examples of strong giant pulses 
detected simultaneously at 1.4 and 2.2~GHz are shown in Figure~\ref{fig:example}.

\section{Giant pulse at 8.3~GHz}\label{sect:res8.3}

\begin{figure}
\begin{center}
    \FigureFile(80mm,60mm){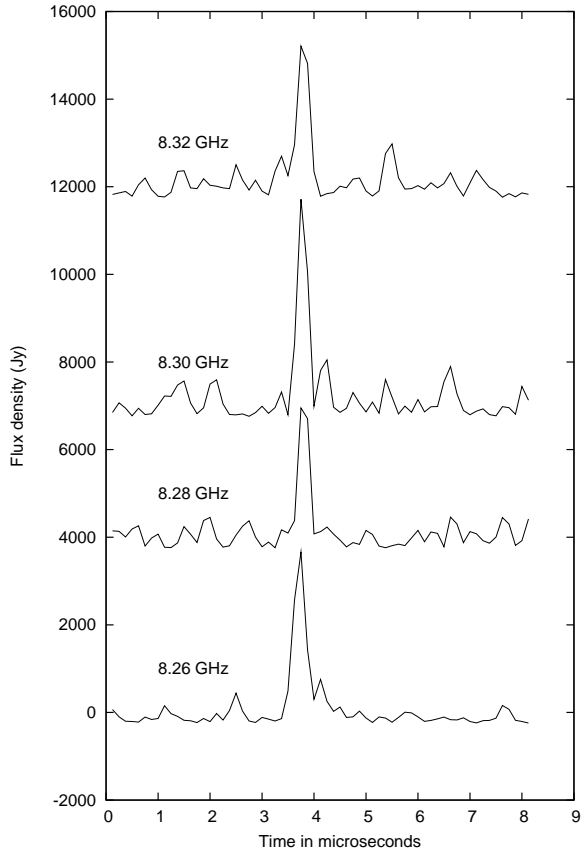}
\end{center}
  \caption{Giant pulse detected at 8.3 GHz.}\label{fig:p8300}
\end{figure}

Only one giant pulse was detected at the observing frequency of 8.3~GHz at a level of $30\sigma$
with a peak flux density of about 10 kJy, nearly equal in all four frequency channels. 
 The pulse profile is shown in Figure~\ref{fig:p8300}.
 The event was not accompanied by notable emision at any of the lower frequency ranges, thus,
proving $\Delta\nu/\nu \le 0.5$ for the radio spectrum of this pulse. On the other hand,
at 8.3 GHz the pulse flux density is constant over about 80~MHz.
The only detection did not allow us to perform a statistical
analysis of GP shapes, widths, and energies as we did for other frequencies (see below). High-frequency
observations of giant pulses from the Crab pulsar were recently carried out by some of us 
with the 100-m Effelsberg radio telescope at 
8.7, 15.1, 22.2~GHz. The first results from the ultra high time resolution data obtained with the Tektronix
oscilloscope were reported by \citet{jessner2008}.

\section{Giant pulses at 1.4 and 2.2 GHz}\label{sect:res1.4-2.2}

Unlike the high frequency of 8.3~GHz where only one GP was detected, we have found 1112 and 631 GP events in all
four frequency channels at the frequency band of 1.4 and 2.2~GHz, respectively (see Table~\ref{tab:freq}).
Below we present a statistical analysis of GP energies and widths, and a study of radio spectra, shapes,
and DM variation.

\subsection{GP width and energy}\label{sect:GPw}

Giant pulses vary significantly in their intensity and width, 
and show a broad variety in their shapes as can be
seen in Figure~\ref{fig:example}. Therefore, it is rather difficult to compose an automated
routine for GP's shape analysis. In this paper we propose two different techniques
for such an analysis: the first one is based on a calculation of an integrated flux density
over varying time boxes in the vicinity of a given GP (this paragraph), and another
method uses analysis of a signal variance (see paragraph~\ref{sub:We}). Let us define the
total width of a giant pulse ($w_{tot}$) as a time interval $\Delta t$ covering the vicinity
of GP which contains $90\%$ of the total integrated flux density of a given GP.
We will calculate the integrated flux density (or pulse energy) as 
$E= \delta t(\sum^n_{i=0} (S_i - \langle S\rangle)),$ where $\delta t$ is sampling time
(in our case $\delta t=0.125\mu s$), $S_i$ are individual counts of reconstructed
flux density. To exclude the noise contribution and calculate the ``net'' 
on-pulse energy, one has to subtract the  mean off-pulse flux density, $\langle S\rangle$,
computed for the off-pulse window<
from flux density value for every sample in on-pulse
window where the required energy is calculated. 
 In this paper we will express pulse energy in units of kJy $\mu$s.
 Following these definitions, we calculated, first, the maximum value for required pulse energy 
separated into two components $E_0^{+}$ and $E_0^{-}$ as
 $E_0^{+} =\delta t( \sum^n_{i=0} (S_{\mathrm{max}+i} - \langle S\rangle))$ and  
$E_0^{-} = \delta t(\sum^n_{i=0} (S_{\mathrm{max}-i} - \langle S\rangle))$.  These quantities
are calculated in the vicinity of a point of maximum flux density of a GP $S_{max}$ 
over $n$ samples in positive and negative time directions. We use $n=256$, 
corresponding to maximum expected pulse width $\Delta t=2n\delta t=64\mu s$.
$E_0^{+}$ and $E_0^{-}$ may differ notably since pulse shape may be not symmetrical
relative its maximum.
With $E_0^{+}$ and $E_0^{-}$ in hand
we calculated trial values $E_k^{+} = \delta t(\sum^k_{i=0} (S_{\mathrm{max}+i} - \langle S\rangle))$ and 
$E_k^{-} = \delta t(\sum^k_{i=0} (S_{\mathrm{max}-i} - \langle S\rangle))$ for every step $k$ in organized loop from $k=0$ to
$n$. On every step $k$ we compared the value of $E_0^{+}$ with the trial value of $E_k^{+}$, and the 
value of $E_0^{-}$ with the value of $E_k^{-}$. When the corresponding difference $|E_0^{\pm} - E_k^{\pm}|$ was 
less than $0.1E_0^{\pm}$, then the loop was terminated and corresponding time $t^{+}$ or $t^{-}$ was
determined as the edge of a GP. Thus, the total GP width $w_\mathrm{tot}$ was defined as
$w_\mathrm{tot} = t^{+} - t^{-}$. Correspondingly, the GP energy $E_\mathrm{p}$ was defined as
$E_\mathrm{p} = E_k^{+} + E_k^{-}$, and the effective width of GP as $w_\mathrm{e} = E_\mathrm{p}/S_\mathrm{max}$. 
Term $w_e$ is traditionally used in pulsar researches, it describes time interval where most
of pulse energy is concentrated, while $w_{tot}$ corresponds to the whole pulse extent.\\

There is a strong constraint on the usage of this approach to analyse GP's shape. As it was already
mentioned, the statistics of a detected signal in our case is close to a $\chi^2$-distribution
with two degrees of freedom. For such a
  signal the root-mean-square deviation is equal to the mean value, i.e.
  the modulation index is equal to 1. Thus, the values of $E_k^{\pm}$
  will manifest notable fluctuations. To keep these fluctuations
  reasonably small one has to select only sufficiently strong pulses.
  We selected a threshold for the peak flux density for such pulses of $50\sigma$
  for the estimated value for maximum pulse width of $w_\mathrm{max} = 64~\mu$s.
  Only 357 and 198 pulses met these requirements at 1.4 and 2.2~GHz, respectively.

\begin{figure*}
    \FigureFile(80mm,60mm){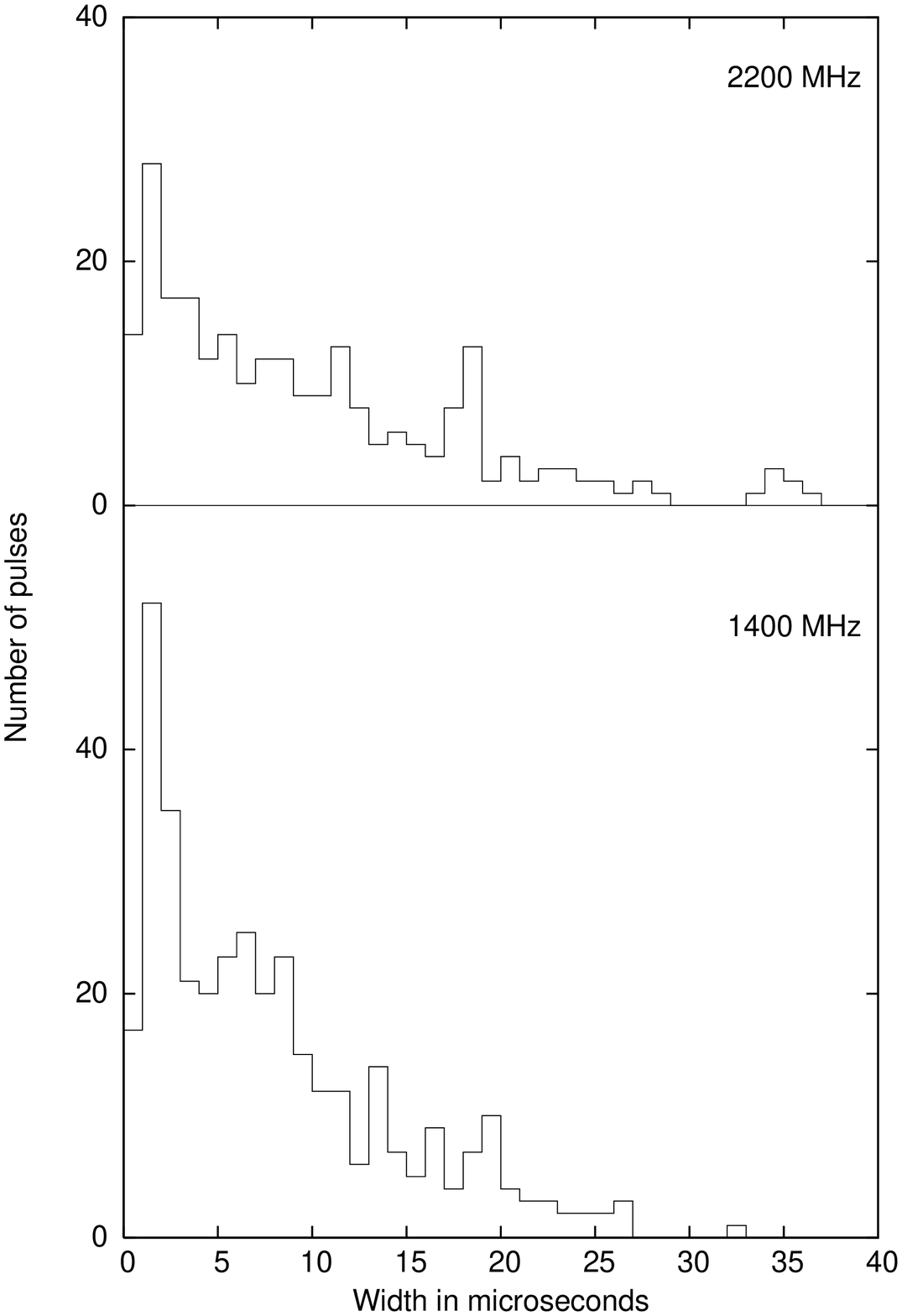}
    \FigureFile(80mm,60mm){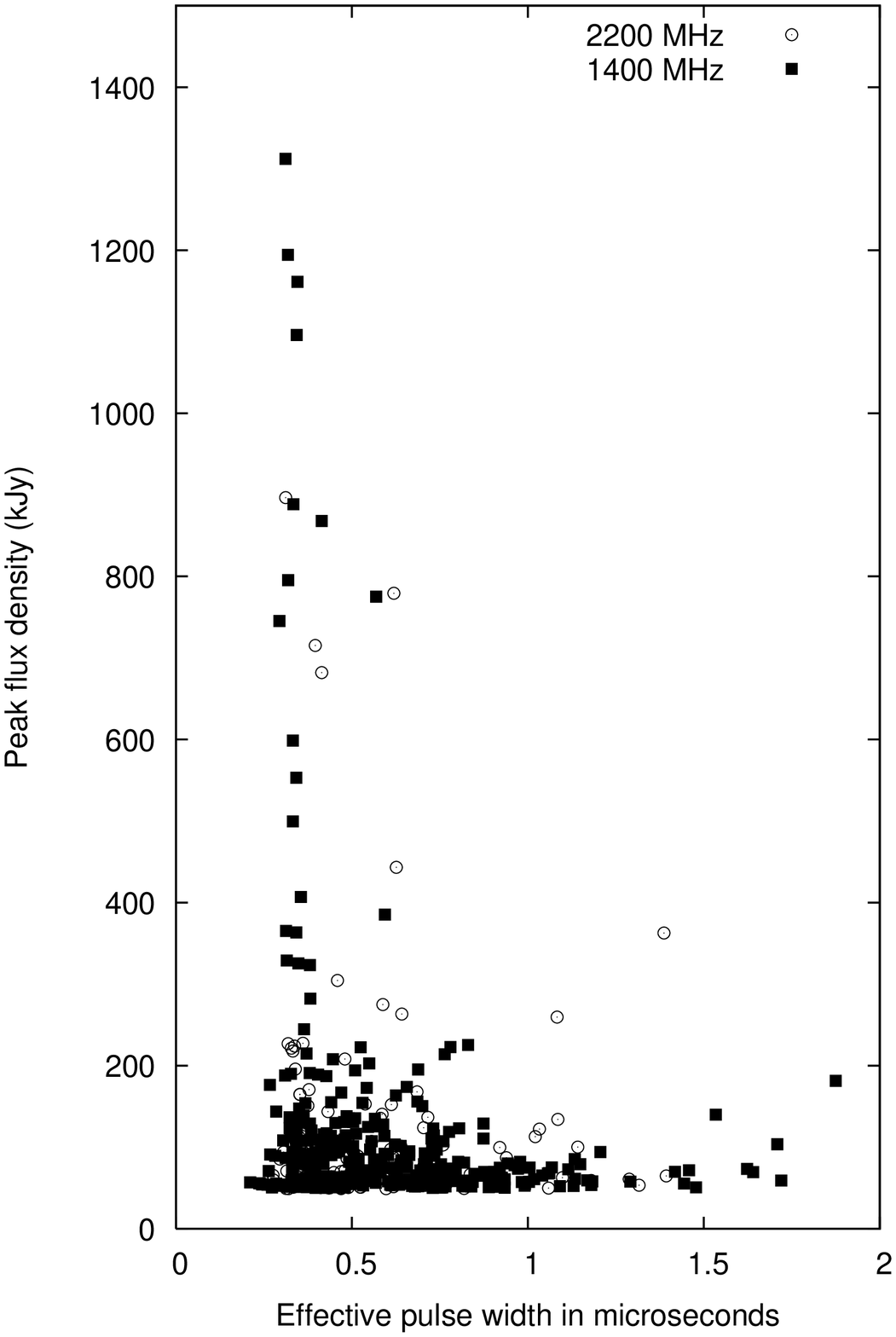}
  \caption{Distribution of the total GP width $w_\mathrm{tot}$ (left), and
  dependence between the peak flux density $S_\mathrm{max}$ and the effective width
  of GP $w_\mathrm{e}$ (right) at 1.4 and 2.2~GHz.}\label{width}
\end{figure*}

  Histograms of the total GP width $w_\mathrm{tot}$  at 1.4 and 2.2~GHz are shown
  in Figure~\ref{width} (left). Both top and bottom histograms are skewed heavily towards
  shorter total widths. Note that only very strong GPs
  were analysed; a weaker population of GPs may have a broader
  width distribution. Right histogram in Figure~\ref{width} represents
  the dependency between
  the maximum flux density, $S_\mathrm{max}$, and the effective pulse width $w_\mathrm{e}$ of a GP.
It is clear that the strongest pulses tend to have shorter durations.
  The same tendency was found by \citet{PopStap2007} at a frequency of 1.2~GHz
on much broader statistics. We did not analyze the cumulative disributions at 1.4 and
2.2~GHz because of rather poor statistics as compared with the data used by
\citet{PopStap2007}, but we will present such an analysis for our data, obtained
at 600 MHz in section~\ref{sect:f600MHz}.

\subsection{Instantaneous GP radio spectra}\label{spectra}

\begin{figure*}[htb]

    \FigureFile(80mm,60mm){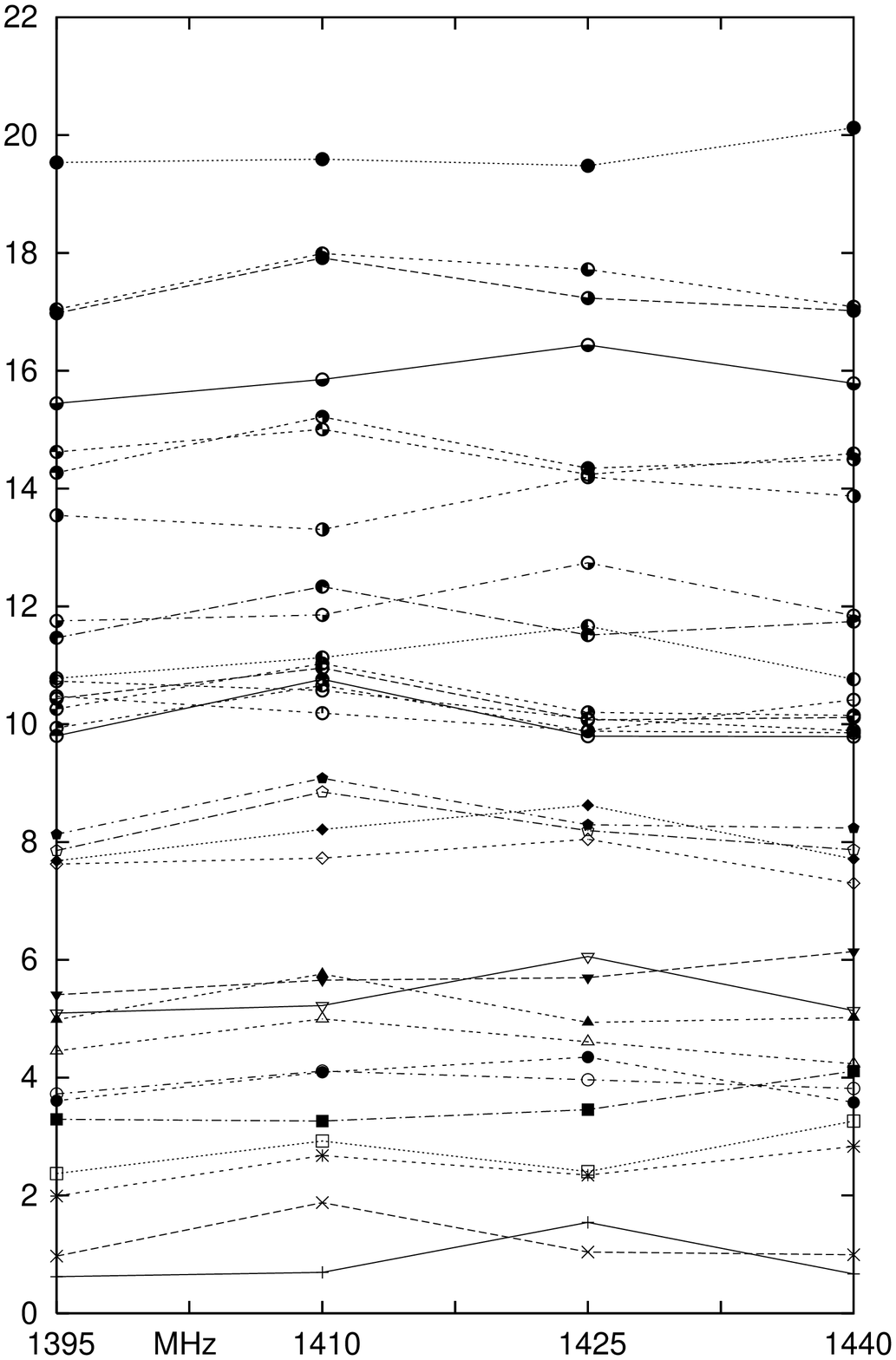}
    \FigureFile(80mm,80mm){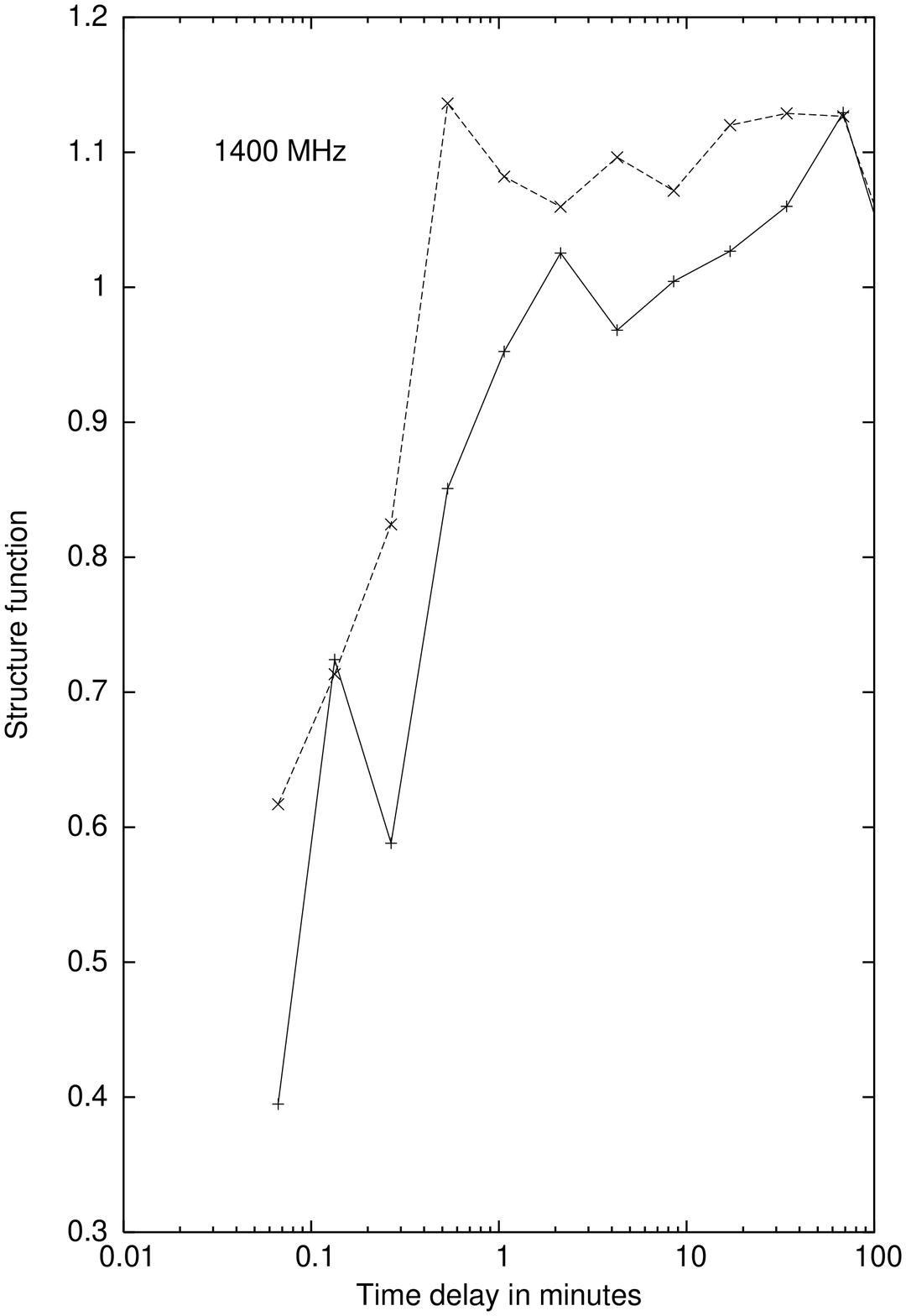}

  \caption{Instant radio spectra of single GP (left)
and their structure functions of time variations for two successive observing days (right)
at 1.4~GHz. Every curve connecting the same symbols in the left figure 
represent the form of instant radio spectrum of a given GP. The location (shift)
of a given spectrum at ordinate axis corresponts to the arrival time of the GP
expressed in minutes from the beginning of the observing set.
}\label{fig:spectra}
\end{figure*}

Given the multifrequency observations with the characteristic frequency setup (see Table~\ref{tab:freq})
we were able to analyze instantaneous radio spectra of detected strong GPs both over large and short
frequency scales. It is important to take into account the effect of interstellar scintillations
when measuring the integrated flux density of a giant pulse in a given receiver band. 
For this we considered results of  another experiment  that was carried out by some of us over the course of our
observations. This experiment was the two-station VLBI observations between the 64-m Kalyazin radio telescope
and 43-m Algonquine radio telescope (Canada) at 2244~MHz using the S2 
recording system in a 16-MHz band (\cite{kondrat_etal}).
In the cross-correlation functions (CCF) of radio spectra of GPs detected at two stations, 
two frequency scales of diffraction modulation of the spectrum at 2244~MHz were found to be
50 and 450~kHz with equal modulation indices \citep{kondrat_etal}.
\citet{cordes2004} found that in the presence of interstellar scintillations the integrated
flux density over the receiver band $B$ will be subject to random fluctuations
which follow a $\chi^2$-distribution with $0.4B/\Delta\nu_d$ degrees of freedom, where $\Delta\nu_d$ 
is the decorrelation bandwidth. In the case of $\Delta\nu_d> 0.2B$ the measured integrated flux density 
is fully modulated by diffractive interstellar scintillations. For the VLBI observations having
$\Delta\nu_d = 450$~kHz at 2.2~GHz and $B=4$~MHz, $\Delta\nu_d>0.1B$ and the modulation
of measured integrated flux densities will be still very significant.
Following the steep power-law frequency dependence of decorrelation bandwidth $\Delta\nu_d$,
$\Delta\nu_d \sim \nu^{4.4}$, for the Kolmogorov spectrum of electron density irregularities 
(see, e.g., Lorimer \& Kramer, 2004), at a frequency of 1.4~GHz, $\Delta\nu_d$ will be about 
5 times less than at 2.2~GHz, and the scintillation modulations will be considerably smoothed.
Therefore, we analysed properties of instantaneous GP radio spectra at short frequency
scales only for the frequency band of 1.4~GHz.
 
The left plot in Figure~\ref{fig:spectra} shows an example of instantaneous GP radio spectra
over 20 minutes of observations at 1.4~GHz. 
Points presented by the same symbols and connected with the same type of  a line
represent the pulse energy measured for a given GP and expressed in arbitrary
units, but with the same scale for an every curve. The every curve is placed at 
the ordinate axis in rough correspondance with arrival time of a given GP.
Arrival time is expressed in minutes from the beginning of a selected
observing interval. The whole figure corresponds to one continuum observing
set. The variations of presented radio spectra with time seem to be random,
but one can see that around the 10th minute
of observations there are four GP spectra similar in appearance.
This fact means that interstellar scintillations do still affect the
measurements. To make a quantitative analysis we calculated
a structure function $C^2(\tau)$ of time variations of instantaneous GP 
A structure function is traditionally used in order to classify 
time variations  of some quantity $Z(t)$ being investigated.
For such a case $C^2(\tau) = \frac{1}{N}\sum_{i=1}^N (Z_i-Z_{i+\tau})^2$.
Then, $C^2(\tau)=0$ with $\tau=0$, and $C^2(\tau)=\sigma^2$ for $\tau > t$ with
$t$ indicating a characteristic time of a variability of $Z(t)$.
Here we transform the definition of $C^2(\tau)$ to apply it
for the analysis of time variations of 4-point instantaneous 
spectra $Y_i^f$ as $C^2(\tau) = \frac{1}{M}\sum_{i=1}^N \sum_{f=1}^4(Y_i^f-Y_{i+\tau}^f)^2$.
Here, M is the number of combinations of radio GP spectra obtained within a
given interval of time delay $\tau$, $N$ is the total number of GPs in this 
test, and $f=1$--4 corresponds to one of four frequency channels listed in Table~\ref{tab:freq}.
The calculated structure functions, $C^2(\tau)$, for two successive observing days
are shown in the right plot of Figure~\ref{fig:spectra}.
It is evident that the structure functions got saturated at time intervals greater
than one minute and decrease rapidly at shorter time delays.
Thus, we do observe the effect of interstellar scintillations at 1.4~GHz that affects
our measurements of radio GP spectra. Nevertheless, even at the shortest time delay 
(less than 10 seconds) between instantaneous GP spectra the structure functions still have a notable 
value of about 0.5 that can be due to an intrinsic modulation of GP spectra
at a short frequency scale of $\Delta\nu/\nu\approx 0.01$.

\begin{figure}[htb]
  \begin{center}
    \FigureFile(80mm,60mm){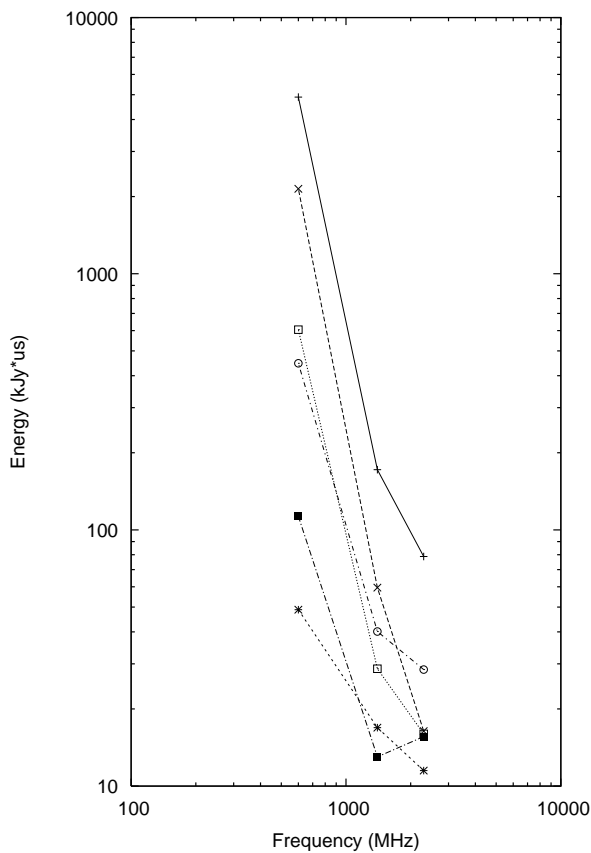}
  \end{center}
  \caption{Instantaneous radio spectra of single GPs in the frequency range of 0.6--2.2~GHz.
Every polygonal line connecting the same symbols represents the spectrum of a selected giant pulse.
}\label{fig:instant}
\end{figure}

Examples of instantaneous radio spectra in the frequency range of 0.6--2.2~GHz for several GPs are presented
in Figure~\ref{fig:instant}.
Every polygonal line connecting the same symbols represents the spectrum of a selected giant pulse.
 GP energies at 1.4 and 2.2~GHz were averaged over 
four frequency channels. There were only 27 strong GPs that 
occurred at all three observing frequencies of 0.6, 1.4, and 2.2~GHz. 
The mean spectral index $\alpha$
is equal to $-2.6$ and $-1.8$ for 0.6--1.4~GHz and 1.4--2.2~GHz, 
respectively\footnote{The spectral index $\alpha$ is defined here as $E_\nu \propto \nu^{\alpha}$,
where $E_\nu$ is the integrated flux density at frequency $\nu$.}
Though our estimates are based on poor statistics of GPs, the tendency for the spectral index
$\alpha$ to flatten at higher frequencies is doubtlessly apparent. The same behaviour
was reported by \citet{Popetal2008} for a frequency range of 600--1650--4850~MHz.

\subsection{DM variations}

\begin{figure*}[htb]             
 
    \FigureFile(80mm,80mm){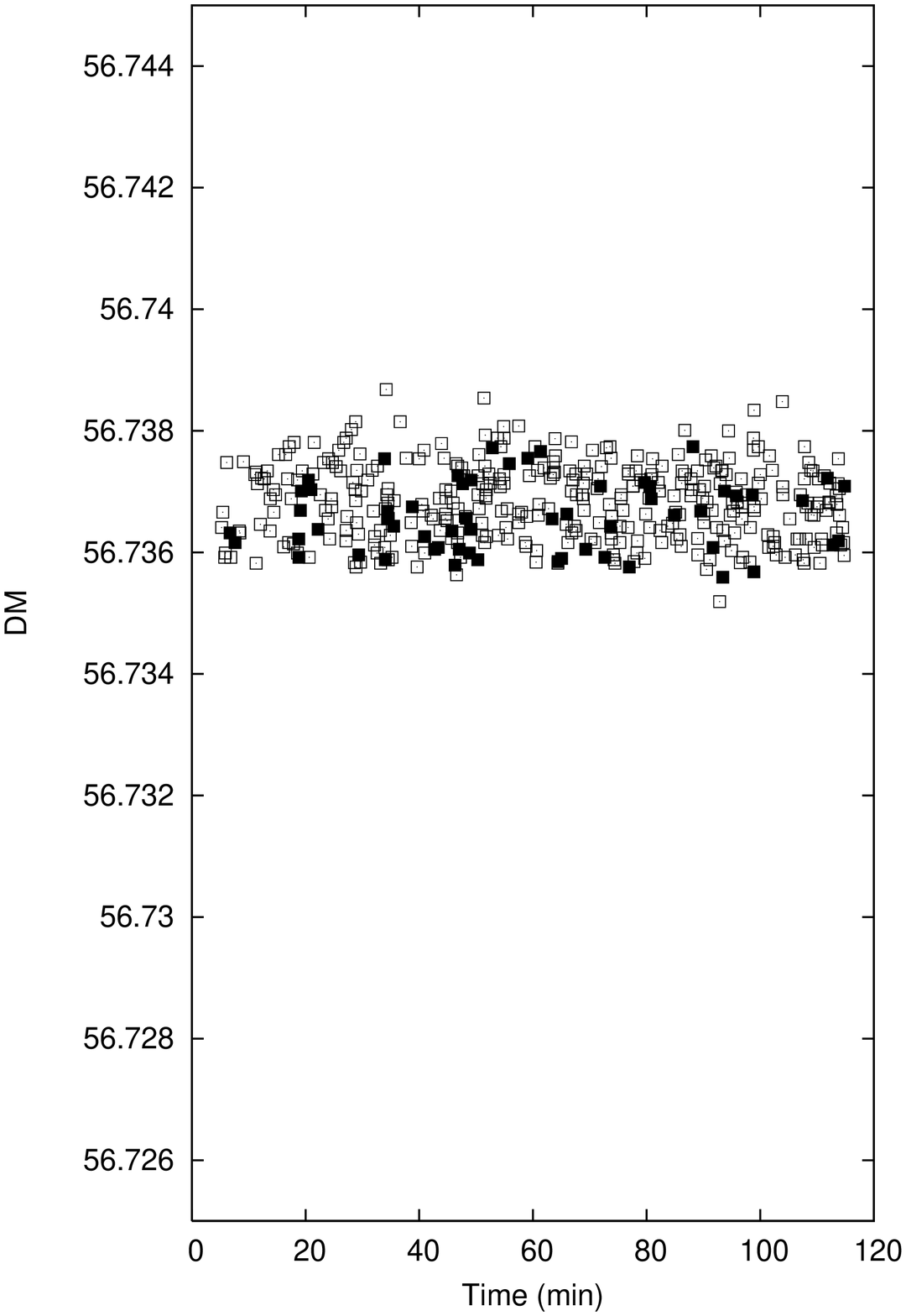}
    \FigureFile(80mm,80mm){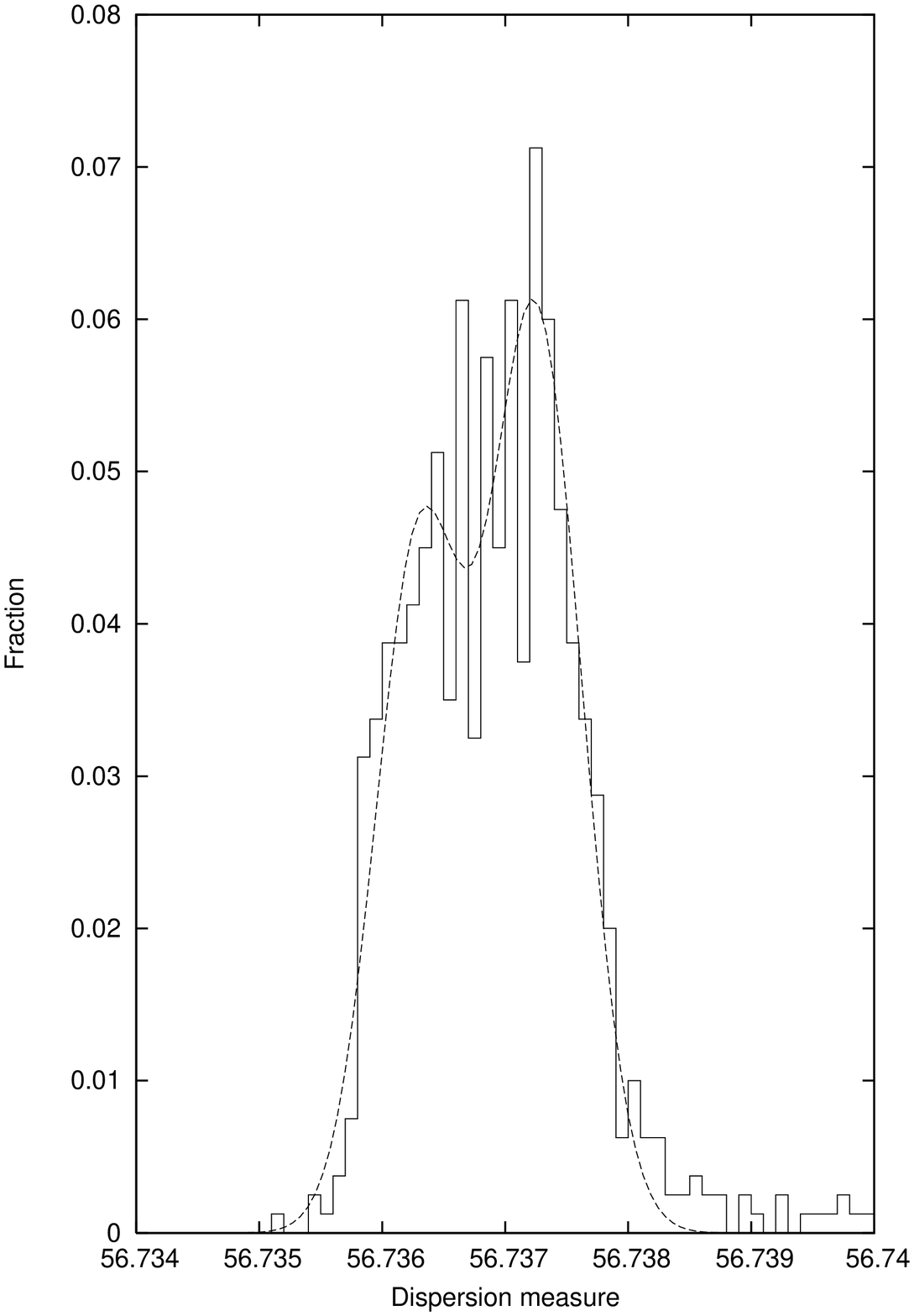}

  \caption{Dispersion measure (DM) variations determined from times of arrival (TOA) of GPs at 600 and 1400~MHz. The left plot
shows the measured DM values for different GPs versus the TOA of the giant pulse. 
 Open squares show GPs that occurred at the longitude of the main pulse, while filled
squares are those that occurred at the longitude of the interpulse. The right plot shows the histogram of DM variations.
The dashed line is the two-Gaussian-component fit with the mean values of 56.7363 and 56.7373~pc cm$^{-3}$. This difference
corresponds to the time shift of about $10~\mu$s at 600~MHz.
}\label{fig:DM}
\end{figure*}

There are 952 GPs out of the total amount of 1112 GP events (or about 85\%) found at 1.4~GHz
that were identified with the GPs detected at 600~MHz. Therefore, we used them to determine
the accurate values of DM by comparing their TOAs at two observing frequencies. The results are
shown in Figure~\ref{fig:DM}.

The left plot of Figure~\ref{fig:DM} represents the dependence of the measured value of DM of GPs versus
their time of arrival. \citet{hanketal2007} pointed out that at high radio frequencies (above 4~GHz) 
DM values of GPs occurring at the longitudes of the interpulse, differ noticeably from DM values
of GPs from the longitude of the main pulse. It is clear that in our observations at lower
frequencies of 0.6 and 1.4~GHz there is no difference between DMs for GPs from different 
longitudes (see Figure~\ref{fig:DM}, left). The right plot of  Figure~\ref{fig:DM} shows the histogram
of the dispersion measures of GPs. One can notice that the DM distribution is likely bimodal and can be
well-fitted by two Gaussians with mean values of 56.7363 and 56.7373~pc cm$^{-3}$. The difference
$\Delta DM=0.001\pm0.0001$ is estimeted with a good formal accuracy. The difference 
corresponds to the time shift of about $10~\mu$s at the observing frequency of 600~MHz. 
As it follows from the left plot in  Figure~\ref{fig:DM} the
splitting can not be attributed to GPs occurred at different longitudes (MP and IP).

\subsection{GP profile shapes}\label{structure}

As will be shown below in section~\ref{subsect:scat600} during our observations the scattering 
contribution was small, and this allowed us to perform the analysis of intrinsic shapes of GPs
detected at the frequency of 2.2~GHz. Indeed, the expected scattering time $\tau_\mathrm{scat}$ 
at this frequency was $\lesssim 200$~ns as extrapolated from the $\tau_\mathrm{scat}$ at 600~MHz, 
which is less than our time resolution of 250~ns (doubled sampling interval).

To study the profiles of individual GPs at the frequency of 2.2~GHz
we used the approach of calculating the autocorrelation functions (ACFs) of GP profiles, 
the technique traditionally used
to study the properties of the microstructure in pulsar radio emission \citep{hank72,cordhank73,Cordes76b}.
However, to avoid the noise spike at zero lag of the ACFs which would mask the presence of any unresolved
components in GP profiles, we calculated CCFs between the profiles of the same GP recorded in different
frequency channels. The cross-correlation function $R_{1,2}(\tau)$ between the profiles $I_1(t)$ and
$I_2(t)$ of the same GP detected in two frequency channels is defined as
$$
R_{1,2}(\tau)=[R_{1,1}(0)R_{2,2}(0)]^{-1/2}\sum_{t=1}^N I_1(t)I_2(t+\tau)~,
$$
where $t$ is the time sample within the GP, $N$
is the total ON-pulse number of samples, and $I_1(t)$, $I_2(t)$ represent ON-pulse
intensities with the OFF-pulse mean level subtracted. In our analysis we chose
the duration of the ON-pulse window of $125~\mu$s, or $N = 1000$ samples. For our 4 frequency
channels at 2.2~GHz there are $C^2_4 = 6$ different frequency combinations listed in the first
column of Table~\ref{tab:delay}.

Due to the effect of interstellar scintillations at this high frequency of 2.2~GHz 
many GPs were not sufficiently simultaneously strong in all frequency channels to provide
a notable correlation coefficient for every pair of GP profiles. Thus, after the calculation
of CCFs for all possible combinations of frequency channels for all 631 GP events found
at 2.2~GHz, we chose only CCFs that have a sufficiently large value of correlation coefficient at zero lag.
For an ACF (or CCF) the root-mean-square deviation ($\sigma_\mathrm{acf}$) near zero lag is 
equal to $\sigma_\mathrm{acf} = 1/\sqrt{N}$. In our case for $N = 1000$, 
$\sigma_\mathrm{acf} \approx 0.03$, and for the subsequent analysis we selected only the CCFs
with the correlation coefficient at zero lag larger than $8\sigma_\mathrm{acf}\approx 0.25$.
There were 121 CCFs selected in this way.

\begin{figure*}[htb]
    \FigureFile(80mm,80mm){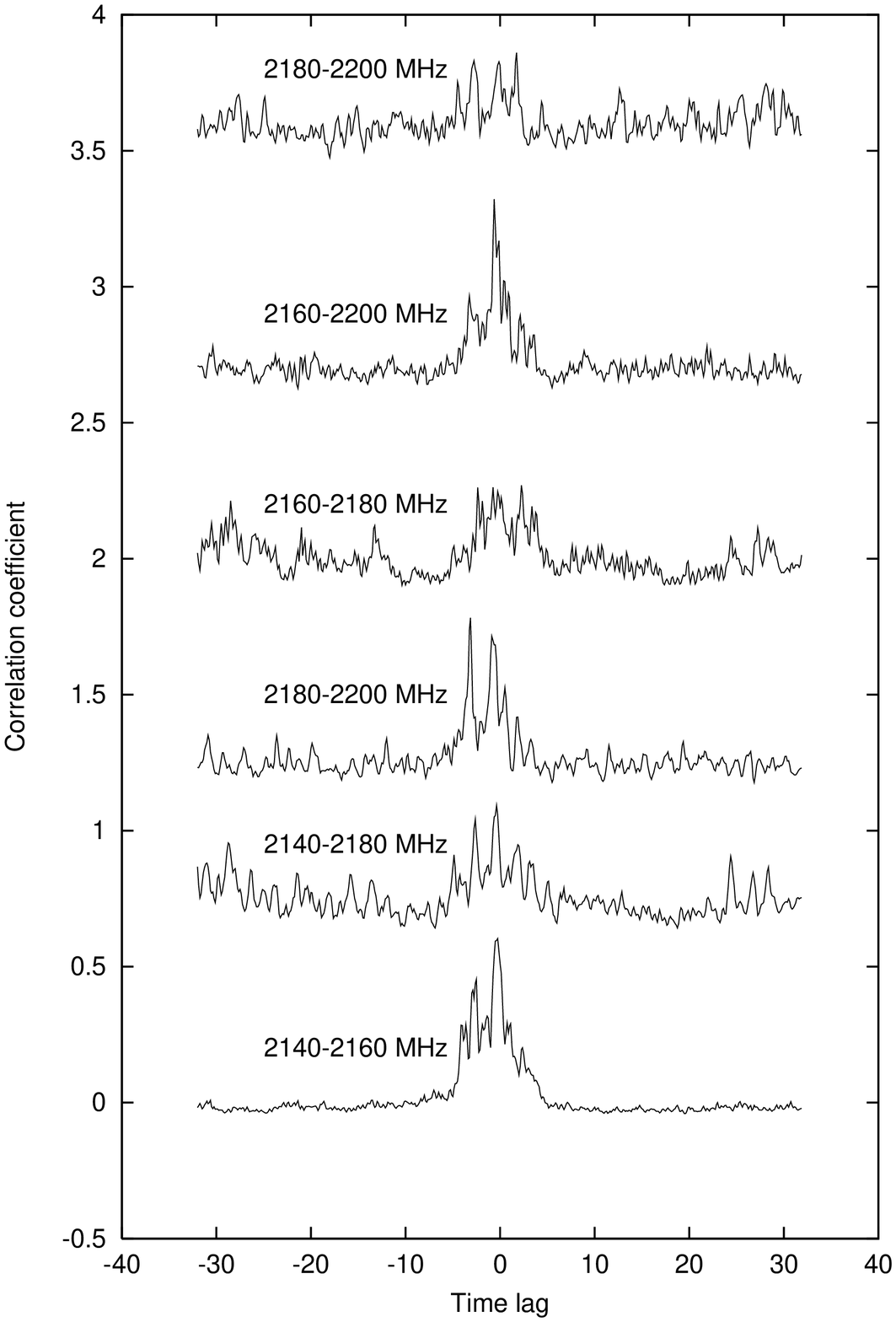}
    \FigureFile(80mm,80mm){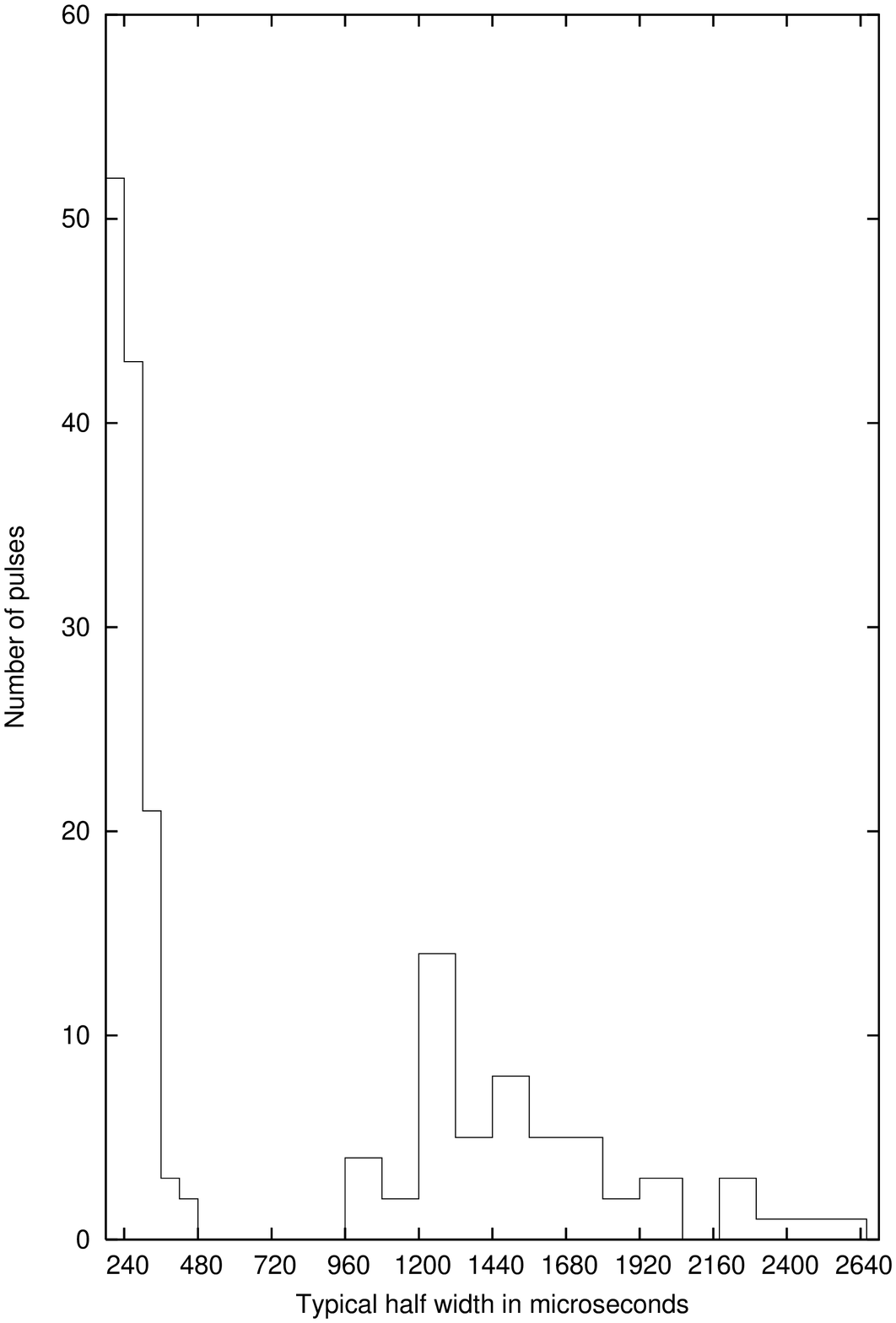}
\caption{Examples of cross-correlation functions (CCFs) between profiles in single frequency channels
for one of the strong GPs at a frequency of 2.2~GHz (left), and the histogram
of the half-widths of the two time scales detected in the CCFs (right).
}
\label{fig:ccf}
\end{figure*}

Figure~\ref{fig:ccf} shows an example of selected CCFs calculated for one of the strong GPs.
One can see the notable correlation between all frequency channels. 
There are two apparent time scales in selected CCFs, namely a short one that occurred as a narrow feature
that can be attributed to the presence of unresolved spikes in the GP structure, and an extended one
with a duration of $\sim 1~\mu$s. The majority of the selected CCFs had the same appearence.
To study the extended time scale feature, we smoothed the CCFs over 20 samples and
selected only those with a correlation coefficient of 0.06 or larger at zero lag. Altogether, 102 of
such CCFs were selected. The right plot of Figure~\ref{fig:ccf} shows the distribution of both
time scales (short and extended) measured at the half-level of the maximum of the corresponding
broad feature. There is a broad jitter of widths of the extended scales in the CCFs from
about $1~\mu$s up to more than $2~\mu$s with the mean value of about 1.3--$1.4~\mu$s. 

In addition to the measurements of the half-widths of the time scales detected in the CCFs, we also 
calculated the time delays of the CCF peaks relative to the expected value for the accepted DM. The
results are listed in the Table~\ref{tab:delay}. One can see that apparent deviations (at a level of 
$>3\sigma$) in GP times of arrival are present only for the frequency channel of 2140~MHz (the lowest
frequency channel at 2.2~GHz). These deviations measured in a 60-MHz frequency range (2140--2200~MHz)
at 2.2~GHz are much smaller than the residual dispersion delays of about $2~\mu$s found in arrival
times of micropulses for the pulsars B0950+08 and B1133+16 at 1.65~GHz over a 16-MHz frequency range
\citep{Popetal2002}.

\begin{table}
  \caption{Extra non-dispersive time delay $\delta t$ of the peak in the selected CCFs between GP profiles in different
frequency channels. These delays represent the residual delays in time of arrival of GPs.
}
  \label{tab:delay}
  \begin{center}
    \begin{tabular}{cc}
      \hline
      Frequency combination & $\delta t$ \\
      (MHz) & ($\mu$s) \\
      \hline
2140--2160 & $-0.20\pm 0.04$ \\
2140--2180 & $-0.17\pm 0.04$ \\
2140--2200 & $-0.26\pm 0.07$ \\
2160--2180 & $\phantom{-}0.05\pm 0.05$ \\
2160--2200 & $-0.06\pm 0.07$ \\
2180--2200 & $-0.01\pm 0.08$ \\
      \hline
    \end{tabular}
  \end{center}
\end{table}

There is also another difference between CCFs computed for GPs in our current analysis
and ACFs/CCFs calculated for the micropulses (see, e.g., \cite{Popetal2002}).
Parameters of the microstructure are explained by the amplitude-modulated noise 
(AMN) model, developed by \citet{Rickett75}. The model explains the pulsar microstructure as an 
envelope of a complex, white Gaussian noise process. Further elaboration of this model was done 
by \citet{Cordes76a} who postulated the origin of the shot noise as a coherent curvature radiation 
from bunched particles. \citet{Cordes76b} showed that ACFs/CCFs for micropulses do follow the
prediction of the AMN model concerning the amplitude of the microstructure component in the
average ACF/CCF. This amplitude is close to 0.5 at zero lag, when properly corrected for a 
signal-to-noise ratio (S/N) and for a time smoothing. Thus, even shortest micropulses do consist 
of a large number of such shots. However, the CCFs calculated for GPs in our analysis
do not obey the prediction of the AMN model, and correlation coefficients at zero lag are
often more than 0.5 level, even without the correction for the finite S/N (see Figure~\ref{fig:ccf}).
Therefore, shortest GP components at 2.2~GHz consist of a small number of individual shots,
or even a single elementary shot.

\section{Giant pulses at 600 MHz}\label{sect:f600MHz} 

Giant pulses at 600~MHz appear more often than at higher frequencies. In more than 5 hours
of observations we found 997 GPs with peak flux density greater than $30\sigma$ --- 849 in main pulse and 148 in interpulse.
The highest peak flux density detected was 2 MJy ($1700\sigma$). 

Owing to unusually low scattering (see section~\ref{subsect:scat600}), we were able to estimate the width of 
the pulses and find the energy distributions separately for pulses of
different widths. Below we will briefly describe our calculation of energies 
and widths, and then discuss the results.

\subsection{Data analysis}

It is well known that GPs from Crab pulsar occur in two narrow
longitude regions of about $6.7^\circ$ each, called main pulse (MP) and interpulse (IP). 
Therefore, we analyzed data only from
three relatively narrow windows: MP, IP and control noise window of the same size 
between MP and IP (with no pulsar emission).

At 600 MHz the search methods differed from those at other frequencies.
Instead of selecting the pulses brighter than fixed threshold, we recorded the ``width'' and ``energy'' in all 
three windows on every period, regardless of the actual presence of giant pulse. 
The properties of GPs themselves were derived from comparison of the whole sets of ``widths'' 
and ``energies'' on MP/IP windows with those of the noise window. 
We believe such method has two main advantages. First of all, it allows us to neglect the unexpected deviations
of noise from Gaussian. Second, it exempts us from selecting pulses by their peak flux density, which can be randomly
biased by scattering. Considering all this, the analysis can be extended to weaker pulses.

\subsubsection{Calculation of width and energy}\label{sub:We}

\begin{figure}[htb]
\begin{center}
\FigureFile(8cm,8cm){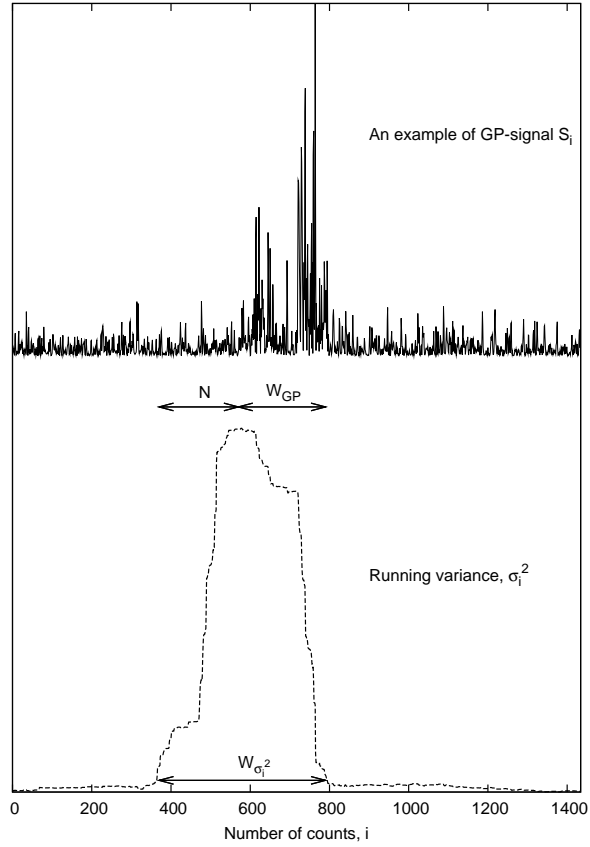}
 \end{center}
  \caption{Example of single pulse detected at 600 MHz, illustrating the technique of pulse width definition.
  Solid line represents the pulse intensity, and dash line shows the runninig variance. 
  See text for more detailed explanation.}
\label{fig:gp_ex600}
\end{figure}

Giant pulses have complex, jagged form, sometimes with several close components (see Figure \ref{fig:gp_ex600}).
Here we considered another technique of GP's width measurements compared with that
proposed in paragraph~\ref{sect:GPw}. 
To find the width of GP, we used the fact that every giant pulse, regardless of its shape and peak 
flux density,  is, in fact, a region with the increased variance of the signal $S$. 
Thus, on every MP/IP/noise window we computed a sequence of variances, 
$\sigma_i^2$ over small floating window:

$$\sigma^2_i=\sum_{k=1}^{N}(S_{i+k}-\bar{S})^2$$~.

The size of the floating window, 
$N$, was comparable to the typical distance between the components (namely 200 samples, i.e. $25\mu s$). 
When such window covered GP, the variance rapidly increased. 
Since the size of the window was about the typical distance between 
the components of GP, the raise and decay were smooth 
and it was easy to measure the width of such outburst which we
designated as $W_{\sigma^2_i}$. It was measured on the level of 5\% of the 
peak maximum (see Figure \ref{fig:gp_ex600}). To obtain the width of GP ($W_{GP}$) we 
subtracted the size of the window $N$ from $W_{\sigma^2_i}$: 

$$W_{GP} = W_{\sigma^2_i} - N~.$$

After that, the signal on corresponding MP/IP or noise window was smoothed with the value of the width. 
The maximum on the smoothed array was regarded as the total energy $E_{tot}$ of an event, and  obtained $W_{GP}$ is understood
as $W_e$. Then we converted derived $W_e$, expressed in samples, to microseconds with $\delta t=0.125\mu s$,
and we scaled $E_{tot}$ to $kJy \cdot \mu s$.

As was mentioned before, such procedure was performed in every period in three different longitude windows. 
On noise window $W_{\sigma_i^2}$ was usually less than the size of floating window $N$, resulting in negative ``GP'' widths. To unify 
the smoothing procedure we assigned all such negative widths the value of 1.

\subsection{GP widths and energies}

 \begin{figure*}[htb]
    \FigureFile(80mm,80mm,){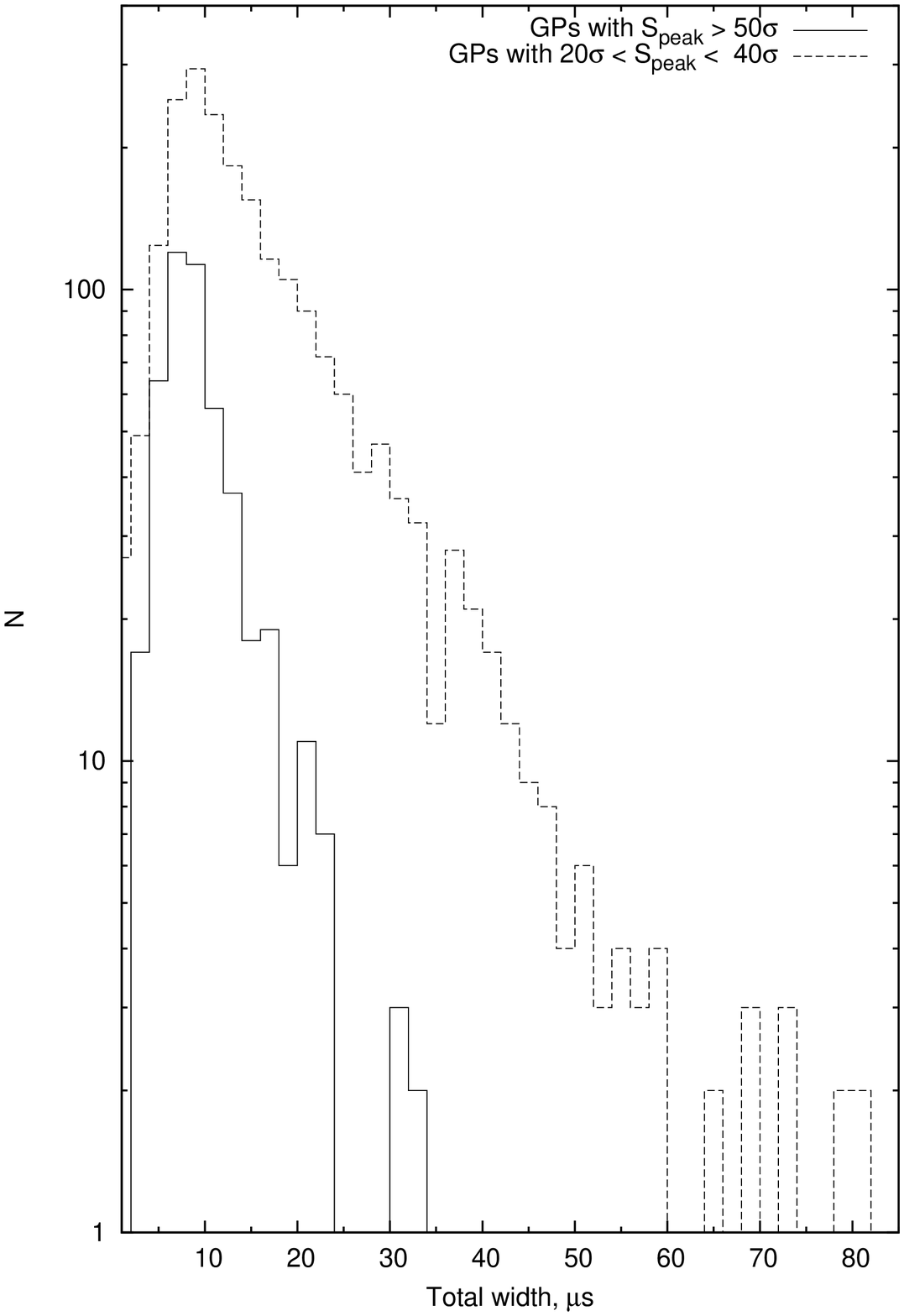}
    \FigureFile(80mm,80mm,){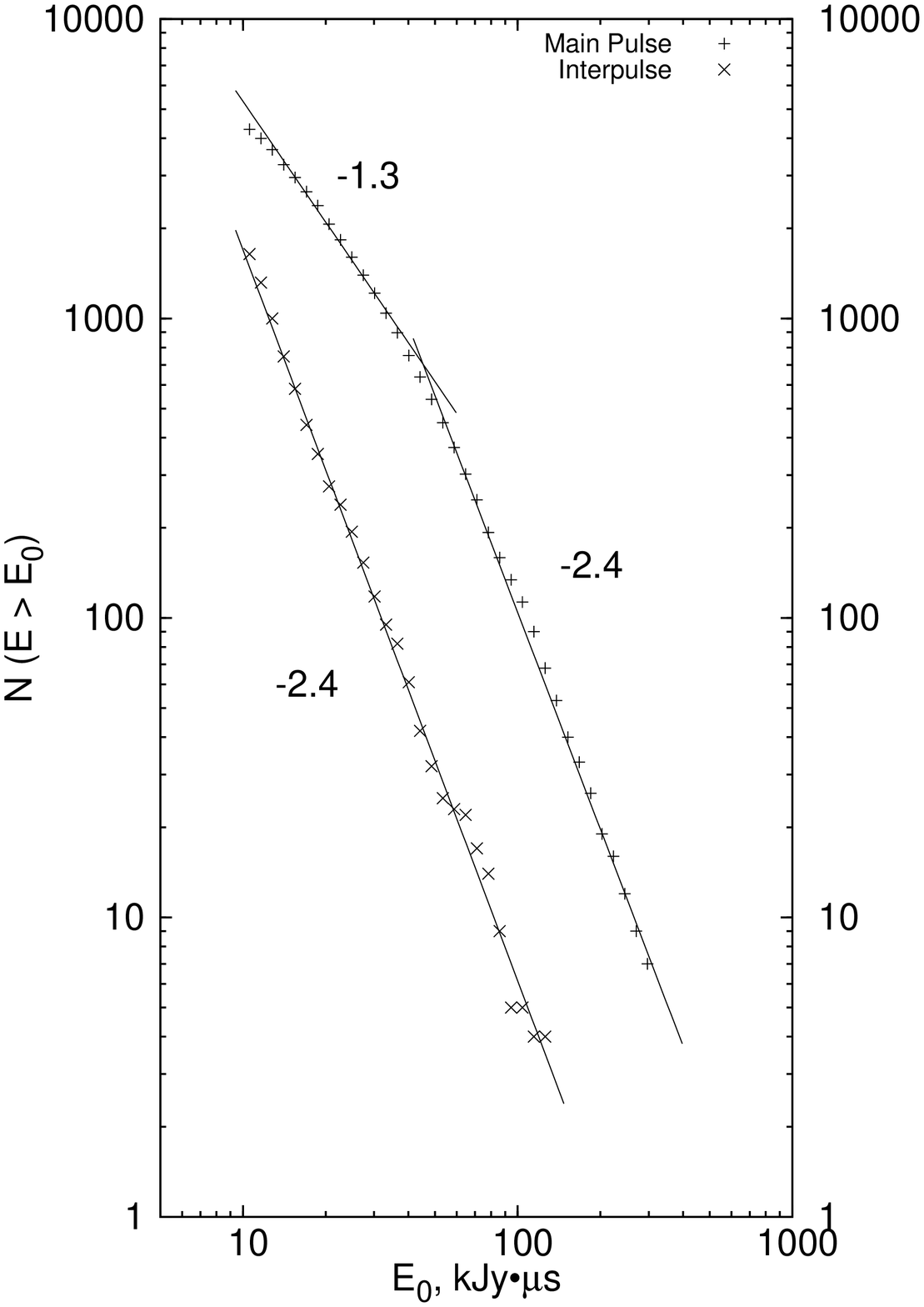}
  \caption{Histogram of total widths (left), and cumulative energy distribution for GPs with widths of 16-24 $\mu$s at 600~MHz (right).
Dash line in the histogram represents GPs with peak flux densities $S_{peak}>50\sigma$ (the strongest GPs),
while solid line is for weaker GPs.}
\label{fig:distr600}
\end{figure*}

As can be seen on Figure \ref{fig:distr600}, the strongest pulses are, on average, narrow, and their width is comparable to the scattering width. The dimmer pulses are usually broader. The same tendency was observed at 1.2 GHz (see section \ref{sect:res1.4-2.2}).

\begin{table}
\caption{The power-law indices $\alpha$ of the cumulative energy distributions of GPs at the frequency of 600~MHz
for different GP widths. Energies of GPs from the longitude of the main pulse have different power-law distributions
for low-energy and high-energy GPs with the clear break energy $E_\mathrm{break}$ (see text) between them. We list
both indices $\alpha_1$ and $\alpha_2$ for this case.}
\label{tab:600_indices}
\begin{center}
 \begin{tabular}{ccccc}
\hline
 Total GP width,  & Interpulse & \multicolumn{3}{c}{Main Pulse}\\
 $\mu$s & $\alpha$ & $\alpha_1$ & $\alpha_2$ & $E_{break}$,\\
  & & &  & Jy$\cdot\mu$s \\
\hline
$< 8 $  & $-1.5$ & \multicolumn{2}{c}{$-1.4$} & --\\
8--16   & $-1.9$ & $-1.2$ & $-1.9$ & 40\\
16--24  & $-2.4$ & $-1.3$ & $-2.4$ & 45\\
$> 24$  & $-3.2$ & $-1.0$ & $-3.8$ & 55\\
\hline
all     & $-2.0$ & $-0.9$ & $-2.2$ & 35\\
\hline
\end{tabular}
\end{center}
\end{table}

The energy distributions for GPs both from the main pulse and the interpulse are well fitted by a power-law,
with the indices depending on the range of widths selected (Table \ref{tab:600_indices}). 
However, for MP GPs wider than 8 $\mu$s, the energy distribution flattens at low energies (see Figure \ref{fig:distr600}), while energy distribution for IP GPs of the same width and energy remains the same.
Thus, the cumulative distribution for GPs from main pulse wider than 8 $\mu$s was fitted with 2 power-law functions: 
$$
N(E>E_0)=\cases{E_0^{\alpha_1}, E_0<E_\mathrm{break};\cr E_0^{\alpha_2}, E_0>E_\mathrm{break}\cr}
$$

Such flattening was previously detected at other frequencies (\cite{Bhat2008,PopStap2007} and references therein). 
The break energy for overall distribution, 35 kJy$\cdot\mu$s, is well consistent with the empiric formula
for the break energy in dependance on the observed frequency, derived by \citet{PopStap2007}:
$$
E_\mathrm{break}=7\nu^{-3.4}~,
$$
where E is in kJy $\cdot\mu$s and $\nu$ in GHz.

 \begin{figure}[htb]
 \begin{center}
    \FigureFile(80mm,80mm){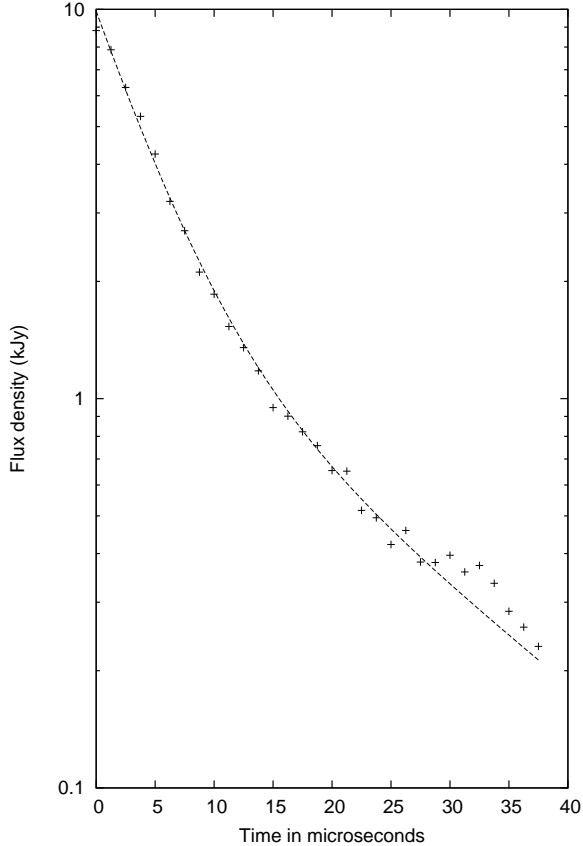}
  \end{center}
  \caption{The tail portion of the average GP profile at 600~MHz folded by summing 1436 GPs
stronger than 50 kJy in peak flux density. The dashed line represents the fit
by two exponents $S = 8.1\exp(-t/\tau_\mathrm{scat,1})+1.8\exp(-t/\tau_\mathrm{scat,2})$ with
the scattering times $\tau_\mathrm{scat,1} = 4.5~\mu$s and $\tau_\mathrm{scat,2} = 17.5~\mu$s.

}\label{fig:scat}
\end{figure}

\subsection{GP scattering}\label{subsect:scat600}

At frequencies below $\sim 1$~GHz pulse shapes are dominated by the scattering broadening 
which decreases with increasing frequency for the Crab pulsar as 
$\tau_\mathrm{scat}\propto f^{-3.5}$ \citep{Popetal2006a}. 
However, the amount of scattering varies significantly with time.
For example, \citet{Popetal2006b} found $\tau_\mathrm{scat}$
to be equal to $45~\mu$s at the frequency of 600~MHz in their observations
with the Kalyazin radio telescope in November, 2003. The measured value  of $\tau_\mathrm{scat}$
was estimated by fitting the exponential function $\exp(-t/\tau_\mathrm{scat})$ to the tail of the average 
GP profile. In our current observations in July, 2005 the scattering broadening was 5--10 times
smaller, and to approximate the average GP profile tail we used
two exponents with the $\tau_\mathrm{scat}$ of 4.5 and $17.5~\mu$s, with short-scale component being
4 times larger in amplitude. The mean GP profile was folded by summing 1436 GPs 
stronger than 50 kJy in peak flux density. They were
preliminary averaged by 10 samples ($1.25~\mu$s), and aligned by their abrupt leading edges. 
The tail portion of the average GP profile at 0.6~GHz and its fit are presented
in Figure~\ref{fig:scat}. Smaller scattering contribution favoured our analysis of intrinsic shapes of GPs 
detected at 2.2~GHz (Section~\ref{structure}).

\section{Conclusions}\label{sect:concl}

Accomplished study of giant pulses from the Crab pulsar from our simultaneous multifrequency 
observations with the Kalyazin radio telescope allowed us to draw a number of conclusions
about the properties of Crab GPs:

\begin{enumerate}

\item Single GPs were often observed simultaneously in the three frequency bands of 0.6, 1.4, and 2.2~GHz, 
confirming the broad band nature of the GP emission. The spectral index $\alpha$ has a tendency to flatten
towards higher frequencies ($\alpha_\mathrm{0.6-1.4} = -2.6$ and $\alpha_\mathrm{1.4-2.2}=-1.8$) as was
also reported before by \citet{Popetal2008}.

\item Instant radio spectra of individual GPs do not follow the single power law in the frequency bands
of 0.6, 1.4, and 2.2~GHz. Many GPs observed both at 0.6 and 2.2~GHz were not detected at 1.4~GHz, thus
proving a notable spectra modulation at a frequency scale of $\Delta\nu/\nu\approx0.5$. At the
frequency of 1.4~GHz in addition to diffraction spectra distortions caused by the wave propagation through
inhomogeneties of the interstellar plasma, we also distinguished the small-scale modulations in radio
spectra at a level of $\Delta\nu/\nu\approx 0.01$ intrinsic to the pulsar.

\item GP profiles at the frequency of 2.2~GHz can be presented by unresolved spikes grouped together
at a time scale of about $1~\mu$s. These unresolved components are well-correlated over 60-MHz
bandwidth ($\Delta\nu/\nu\approx0.03$). For comparison, \citet{Hankins2000} did not find the correlation
for the unresolved structure of GPs between 4535 and 4985~MHz (i.e., $\Delta\nu/\nu\approx 0.1$).

\item Cross-correlation functions of GP profiles in different frequency channels do not obey the 
prediction of the AMN model \citep{Rickett75} for the microstructure of the pulsar radio emission.
Thus, unresolved components represent a small number or even the single elementary emitter.

\item The extra non-dispersive time delays of about $0.2~\mu$s found in arrival times of unresolved components 
of GP structure at 2.2~GHz are not notable and 10 times smaller than the residual dispersion delays
of about $2~\mu$s found in TOAs of micropulses for the pulsars B0950+08 and B1133+16 at 1.65~GHz \citep{Popetal2002}.

\item Strongest GPs tend to have shorter durations. Same results were reported by 
\citet{PopStap2007} and by \citet{Bhat2008} at  frequencies of 1.2--1.4~GHz.

\item The distribution of dispersion measures of GPs determined from their TOAs at 0.6 and 1.4~GHz
is likely bimodal and can be well-fitted by two Gaussians with mean values of 56.7363 and 56.7373~pc cm$^{-3}$.

\item The cumulative distribution of GP energies at 600~MHz depends on the effective width of GPs. This
distribution obtained only for GPs at the longitude of the main pulse, has the break and can be represented 
by two power-law functions with indices of -2.0 -- -0.9 at high and low energies respectively.  The value of the energy $E_\mathrm{break}$, where
this break occurred, is equal to 35 kJy$\cdot \mu s$, and it fits to the empiric 
formula $E_\mathrm{break}(\mathrm{kJy \cdot \mu s}) \approx 7\nu^{-3.4}$ with $\nu$ in GHz derived by \citet{PopStap2007}.

\end{enumerate}

\end{document}